\documentclass{WileyASNA-v1}

\articletype{Article Type}%
\begin{document}

\title{On the Connection between Radiative Outbursts and Timing Irregularities in Magnetars}

\author[1]{C.-P. Hu*}

\author[1]{C.-Y. Ng}

\authormark{Hu \& Ng}

\address[1]{\orgdiv{Department of Physics}, \orgname{The University of Hong Kong}, \orgaddress{Pokfulam Road}, \country{Hong Kong}}
\corres{*Chin-Ping Hu \email{x1916053@gmail.com}}

\abstract{Magnetars are strongly magnetized pulsars and they occasionally show violent radiative outbursts. They also often exhibit glitches that are sudden changes in the spin frequency. It was found that some glitches were associated with outbursts but their connection remains unclear. We present a systematic study to identify possible correlations between them. We find that the glitch size of magnetars likely shows a bimodal distribution, different from the distribution of the Vela-like recurrent glitches but consistent with the high-end of that of normal pulsars. A glitch is likely a necessary condition for an outburst but not a sufficient condition because only 30\% of glitches were associated with outbursts. In the outburst cases, the glitches tend to induce larger frequency changes compared to those unassociated ones.We argue that a larger glitch is more likely to trigger the outburst mechanism, either by reconfiguration of the magnetosphere or deformation of the crust. A more frequent and deeper monitoring of magnetars is necessary for further investigation of their connection. }



\maketitle

\section{Introduction}\label{introduction}

Pulsars are one of the most precise clocks in the Universe. They are powered by rotational energy and show periodic signals with gradual spin-down. Two types of timing irregularities, timing noise and glitches, were commonly found in young pulsars. They provide hints of the stellar interior and its interaction with the magnetosphere. Timing noises have several forms and their mechanisms remain unclear \citep[see, e.g., ][]{HobbsLK2010, LyneHK2010}. A glitch is a sudden increase of the spin frequency, and it is often followed by a recovery \citep{McCullochHR1983, EspinozaLS2011}. A statistical analysis shows that the glitch size is bimodally distributed, which could indicate different triggering mechanisms \citep{EspinozaLS2011}. Several theoretical interpretations have been proposed, such as rearrangement of the crust shape triggered by starquakes \citep{BaymPP1969, BaymP1971} and a catastrophic break down of vortex pinning in the superfluid component \citep{AlparPA1984, AndersonI1975}. 

Magnetars are a special class of pulsars that contain extremely high magnetic fields \citep[see review by ][]{KaspiB2017}. The most remarkable features of them are the short-term bursts with a time scale of seconds and long-term outbursts with a time scale from months to years. They usually have thermal luminosities higher than that inferred from the spin-down and hence are believed to be powered by the decay of the magnetic field \citep{DuncanT1992}. The triggering mechanisms of the burst and outburst remain clouded with controversy. A burst could be triggered internally such as instability of the core and cracking of the crust \citep{ThompsonD1995, ThompsonD2001}, or externally like a sudden reconnection of a twisted magnetosphere \citep{Lyutikov2003, ParfreyBH2013}. An outburst, generally accompanied by an intensive burst epoch \citep{WoodsKF2007}, could be powered by gradually untwisting of the magnetosphere \citep{Beloborodov2009, ThompsonLK2002}. Observations of several magnetars showed that additional hotspots, which are originated from the bombardment by particles accelerated in the magnetosphere, shrank gradually during the tail of the outburst and hence supporting this model \citep{BeloborodovL2016}. 

Magnetars also show glitches frequently. Five bright magnetars, 1E~1841$-$045, 1RXS~J170849.0$-$400910, 1E~2259+586, 4U~0142+61, and 1E~1048.1$-$5937, have been monitored with the \textit{Rossi X-ray Timing Explorer} (\textit{RXTE}) between 1996 and 2012, and 17 glitches/timing anomalies were observed \citep{DibK2014}. Their fractional glitch sizes ($\Delta\nu/\nu$) are huge but their absolute size ($\Delta\nu$) spreads over a wide range with much lower values than those of Vela-like pulsars \citep{EspinozaLS2011}. Moreover, all the outbursts were accompanied by glitches, but not vice versa \citep{DibK2014}. Timing anomaly and radiative outburst are believed to have some connection because they share several common origins. However, it remains unclear if glitches associated with outburst have any distinct properties compared to others. This motivates us to examine the differences between these two types of glitches with an extended database. 

We describe the current glitch sample of magnetars in section \ref{sample}. The statistic of the glitch size and its correlation with physical properties are shown in section \ref{corr}. We discuss the possible connection between the radiative outbursts and the glitches in Section \ref{discussion}. We then summarize our work and propose future prospects in Section \ref{summary}.

\section{Glitch Sample}\label{sample}

We select the magnetar sample from the McGill Online Magnetar Catalog\footnote{\url{http://www.physics.mcgill.ca/~pulsar/magnetar/main.html}}, which categorizes 23 confirmed magnetars \citep{OlausenK2014}. The glitch sample from five bright magnetars is adopted from \citet{DibK2014} with a few updates \citep{ArchibaldKN2015, ArchibaldKS2017}. For other magnetars, we obtain glitches in CXOU J164710.2$-$455216, 1E~1547.0$-$5408, SGR~J1745$-$2900, and Swift~J1822.3$-$1606. We also include high magnetic-field rotation-powered pulsars (RPPs) that have magnetic field strengths of $B=10^{13}$--$10^{14}$\,G. They are believed to be the transitional class of pulsars bridging magnetars and canonical RPPs \citep[see, e.g.,][]{KaspiM2005, NgK2011, HuNT2017}. We only choose PSRs\,J1846$-$0258 and J1119$-$6127 because they are confirmed to show magnetar-like behaviors \citep{GavriilGG2008, ArchibaldKT2016, GogusLK2016}. Several magnetars show anti-glitches where the spin frequencies jump to lower values \citep{GavriilDK2011, ArchibaldKN2013, SasmazAG2014}. We do not include them in the following analysis since their mechanism could be dramatically different from canonical glitches \citep{ArchibaldKN2013,DibK2014,PintoreBM2016}. Some anti-glitches are likely over-recovery from spin-up glitches and we include their spin-up measurement in the analysis \citep{GavriilDK2011, ArchibaldKS2017}. All the sampled glitches are listed in Table \ref{glitch_table}. 

\begin{center}
\begin{table*}[t]%
\caption{Glitches in Magnetars and their Assoication with Radiative Outbursts.\label{glitch_table}}
\centering
\begin{tabular*}{500pt}{@{\extracolsep\fill}lccccc@{\extracolsep\fill}}
\toprule
\textbf{Name} & \textbf{Time}  & \textbf{$\Delta\nu$}  & \textbf{$\Delta\dot{\nu}$}  & \textbf{Outburst} & \textbf{References}  \\
 & (MJD) & (Hz) & (Hz\,s$^{-1}$) & & \\
\midrule
1E 1841 & 52453 & $2.9(1)\times10^{-7}$ & $-1.3(1)\times10^{-14}$  & N & \citet{DibKG2008}\\
 & 52997 & $2.08(4)\times10^{-7}$ & $4(3)\times10^{-16}$ & N & \citet{DibKG2008}\\
 & 53823 & $1.17(9)\times10^{-7}$ & $2(1)\times10^{-15}$ & N & \citet{DibKG2008}\\
 & 54304 & $4.6(3)\times10^{-7}$ & $-2(1)\times10^{-14}$ & N & \citet{DibKG2008}\\
 & 55596 & $8.2(7)\times10^{-8}$ & $4(1)\times10^{-15}$ & N & \citet{DibK2014}\\
1RXS~J1708 & 51445 & $5.1(3)\times10^{-8}$ & $-8(4)\times10^{-16}$ & N & \citet{KaspiLC2000}\\
 & 52016 & $3.6(3)\times10^{-7}$ & $-1.1(2)\times10^{-15}$ & N & \citet{KaspiGW2003}\\
 & 52990 & $2.8(4)\times10^{-8}$ & N/A & N & \citet{DibKG2008}\\
 & 53366 & $5.5(8)\times10^{-8}$ & $-2(1)\times10^{-15}$ & N & \citet{DibK2014}\\
 & 53549 & $2.5(9)\times10^{-7}$ & $-2(2)\times10^{-15}$ & N & \citet{DibK2014}\\
 & 53636 & $6.7(3)\times10^{-8}$ & $6(5)\times10^{-15}$ & N & \citet{DibKG2008}\\
 & 55517 & $9.4(5)\times10^{-8}$ & $1.4(4)\times10^{-15}$ & N & \citet{DibK2014}\\
1E~2259 & 52443 & $5.0(1)\times10^{-7}$ & $2.2(3)\times10^{-16}$ & Y & \citet{WoodsKT2004}\\
 & 54184 & $1.261(4)\times10^{-7}$ & $-6(2)\times10^{-16}$ & N & \citet{DibKG2008}\\
 & 53750 & $4.4(5)\times10^{-9}$ & N/A & N & \citet{IcdemBI2012}\\
 & 54040 & $>1.6\times10^{-6}$ & N/A & N & \citet{IcdemBI2012}\\
 & 54856 & $1.2(3)\times10^{-8}$ & $2.3(1.6)\times10^{-16}$ & Y & \citet{IcdemBI2012, DibK2014}\\
 & 56125 & $3.6(7)\times10^{-8}$ & $2.6(2)\times10^{-14}$ & Y & \citet{ArchibaldKN2013}\\
4U~0142 & 51251 & $7.4(7)\times10^{-8}$ & $-2.4(3)\times10^{-16}$ & N & \citet{MoriiKS2005}\\
 & & & & & \citet{DibKG2007}\\
 & 53809 & $2.0(4)\times10^{-7}$ & $-3(1)\times10^{-16}$ & Y$^a$ & \citet{GavriilDK2011}\\
 & 55771 & $5.11(4)\times10^{-7}$ & $-6(4)\times10^{-14}$ & N & \citet{DibK2014}\\
 & 57081 & $5.1(5)\times10^{-8}$ & N/A & Y$^b$ & \citet{ArchibaldKS2017}\\ 
1E~1048 & 52218 & $\sim1\times10^{-7}$ & N/A & Y & \citet{DibKG2009}\\
 & 52386 & $4.5(1)\times10^{-7}$ & $-4(1)\times10^{-14}$ & Y & \citet{DibKG2009}\\
 & 54185 & $2.52(3)\times10^{-6}$ & $-6(4)\times10^{-14}$ & Y & \citet{DibKG2009}\\
 & 55926$^c$ & N/A & $\sim4\times10^{-14}$ & Y & \citet{ArchibaldKN2015}\\
CXOU~J1647 & 53968 & $6.1(3)\times10^{-7}$ & N/A & Y & \citet{IsraelCD2007}\\
1E~1547 & 54853 & $9(7)\times10^{-7}$ & $7.7(8)\times10^{-12}$ & Y & \citet{KuiperHH2012}\\
SGR~1745 & 56450 & $<3\times10{-7}$ & $-5.5(1)\times10^{-13}$ & N & \citet{KaspiAB2014}\\
Swift~J1822 & 56756 & $2.7(1)\times10^{-8}$ & N/A & Y & \citet{ScholzKC2014}\\
PSR~J1119 & 51398 & $1.1(1)\times10^{-8}$ & $-9(1)\times10^{-16}$ & N & \citet{JanssenS2006}\\
 & 57596 & $1.40(2)\times10^{-5}$ & $-1.9(2)\times10^{-12}$ & Y & \citet{ArchibaldKT2016}\\
PSR~J1846 & 53883 & $1.2(4)\times10^{-5}$ & $-2.7(1)\times10^{-13}$ & Y & \citet{LivingstoneKG2010}\\
\bottomrule
\end{tabular*}
\begin{tablenotes}
\item[\textit{a}] ~~Followed by a spin-down glitch of $\Delta\nu=-1.27(2)\times10^{-8}$ and a short-term, limited flux increase.
\item[\textit{b}] ~~Followed by a spin-down glitch of $\Delta\nu=-3.7(1)\times10^{-8}$ and a short-term, limited flux increase.
\item[\textit{c}] ~~A change in torque, but unlikely to be a glitch.
\end{tablenotes}
\end{table*}
\end{center}

\section{Analysis Results}\label{corr}
\begin{figure*}
\begin{minipage}{0.49\linewidth}
\includegraphics[width=0.95\textwidth]{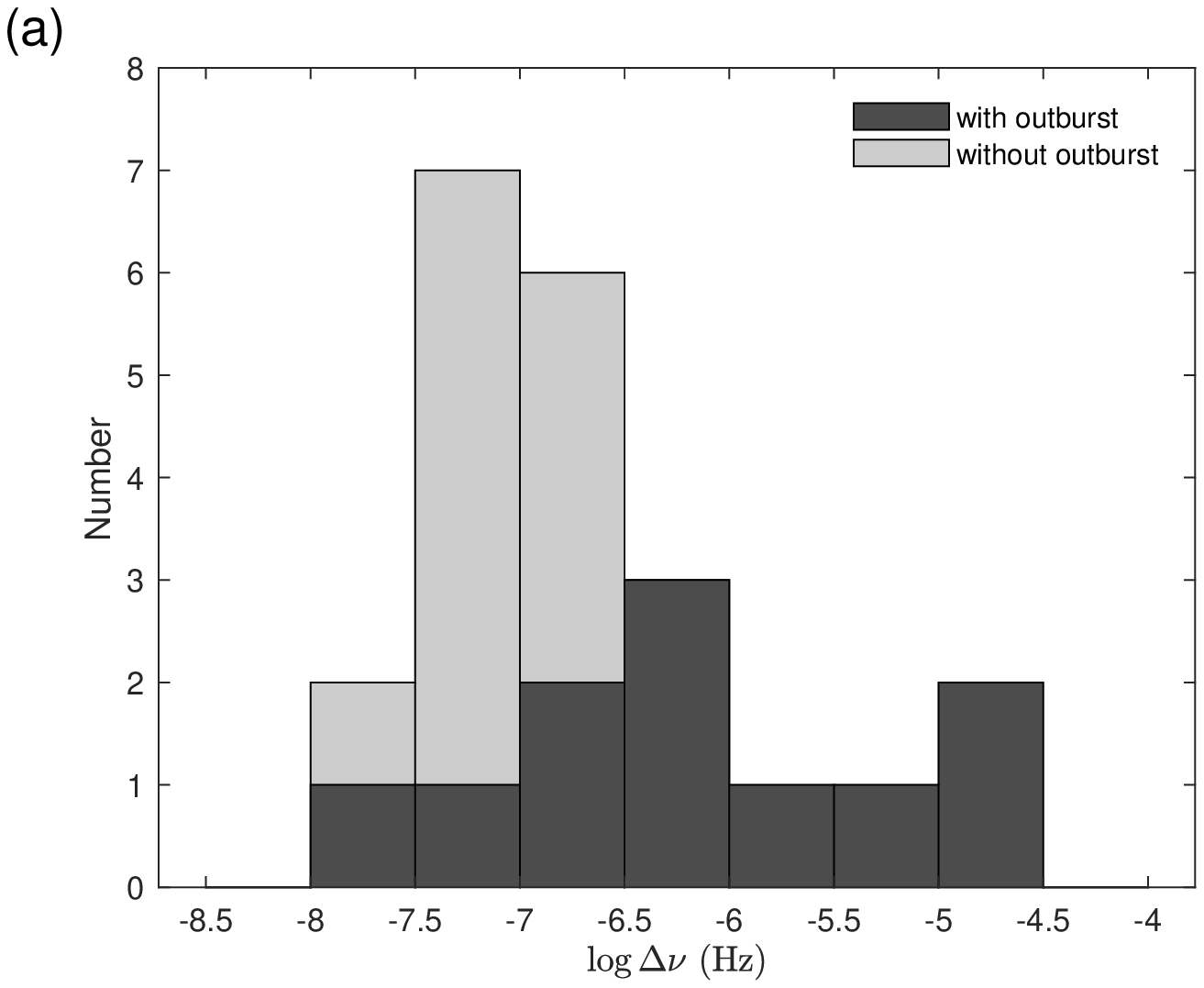}
\end{minipage}
\begin{minipage}{0.49\linewidth}
\includegraphics[width=0.95\textwidth]{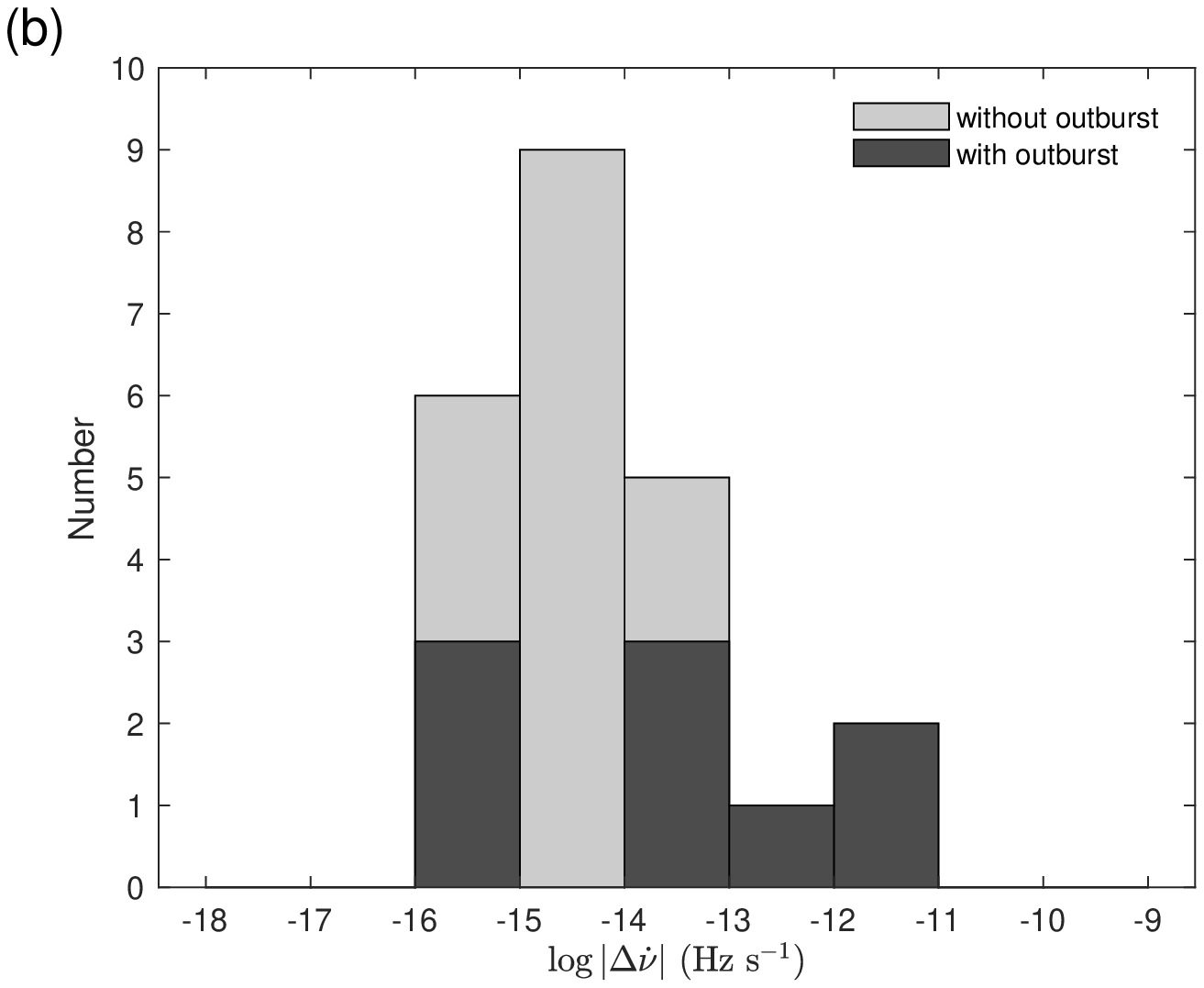}
\end{minipage}
\caption{Histogram of the glitch size in (a) $\Delta\nu$ and (b) $|\Delta\dot{\nu}|$ of magnetars. The glitches associated with outburst are marked with dark gray, while other ones are marked with light gray. }
\label{df_df1_histogram}
\end{figure*}

We first investigate the glitch size distribution of magnetars. Figure \ref{df_df1_histogram} shows the histogram of jumps in frequency ($\Delta\nu$) and frequency derivative ($\Delta\dot{\nu}$). Glitches without outburst have Gaussian-like distributions in $10^{-8}$\,Hz$<\Delta\nu<10^{-6}$\,Hz and $10^{-16}$\,Hz\,s$^{-1}<\Delta|\dot{\nu}|<10^{-12}$\,Hz\,s$^{-1}$. They are consistent with the high-end of the glitch distribution of the major pulsar population \citep[see][]{EspinozaLS2011} although they are mainly observed from limited sample. On the other hand, glitches with outbursts have a much wider distribution in $\Delta\nu$ but have no significantly different distribution in $\Delta|\dot{\nu}|$. They occupied the saddle between the major pulsar population and the Vela-like pulsars. 

\begin{figure*}
\begin{minipage}{0.49\linewidth}
\includegraphics[width=0.95\textwidth]{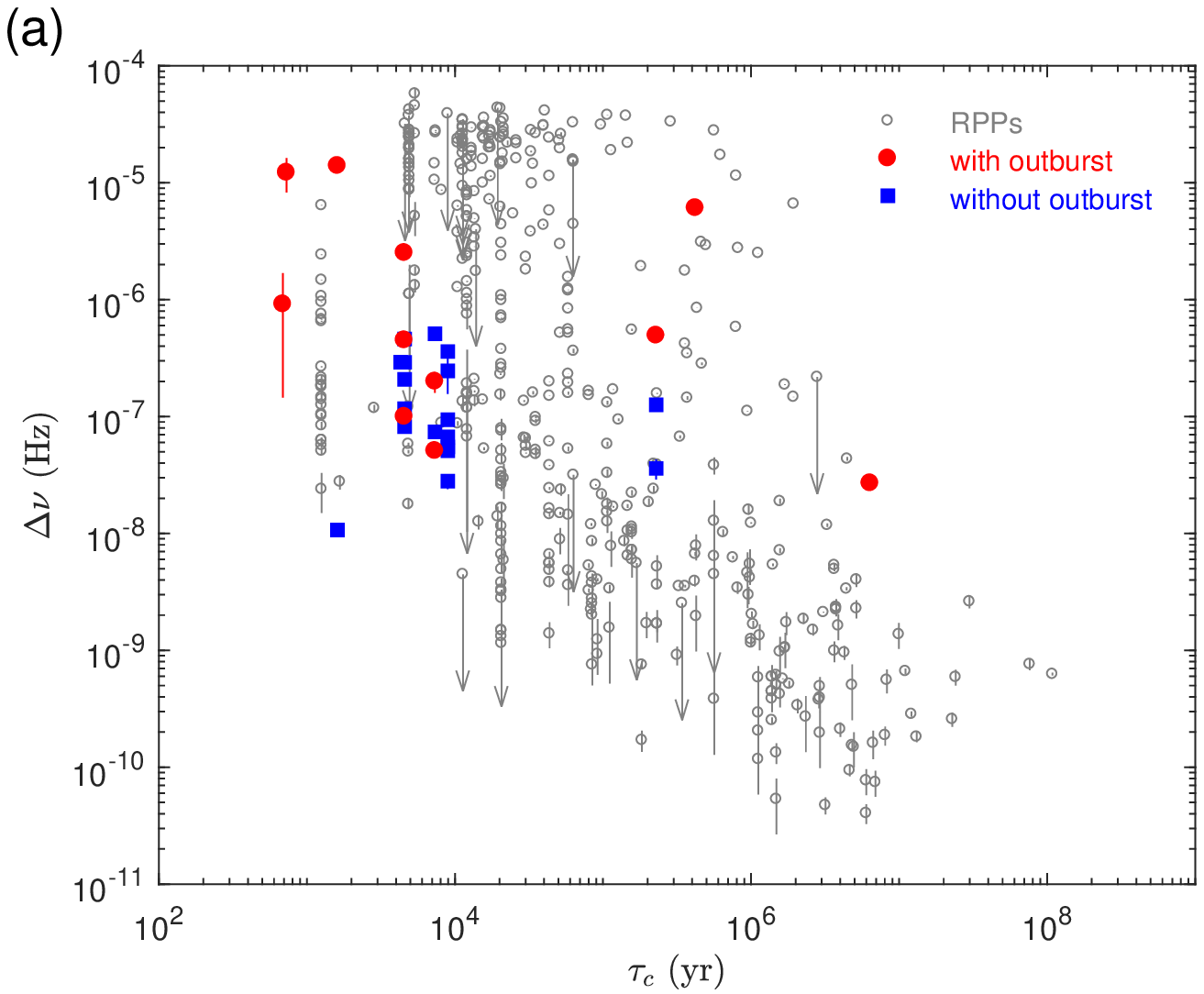}
\end{minipage}
\begin{minipage}{0.49\linewidth}
\includegraphics[width=0.95\textwidth]{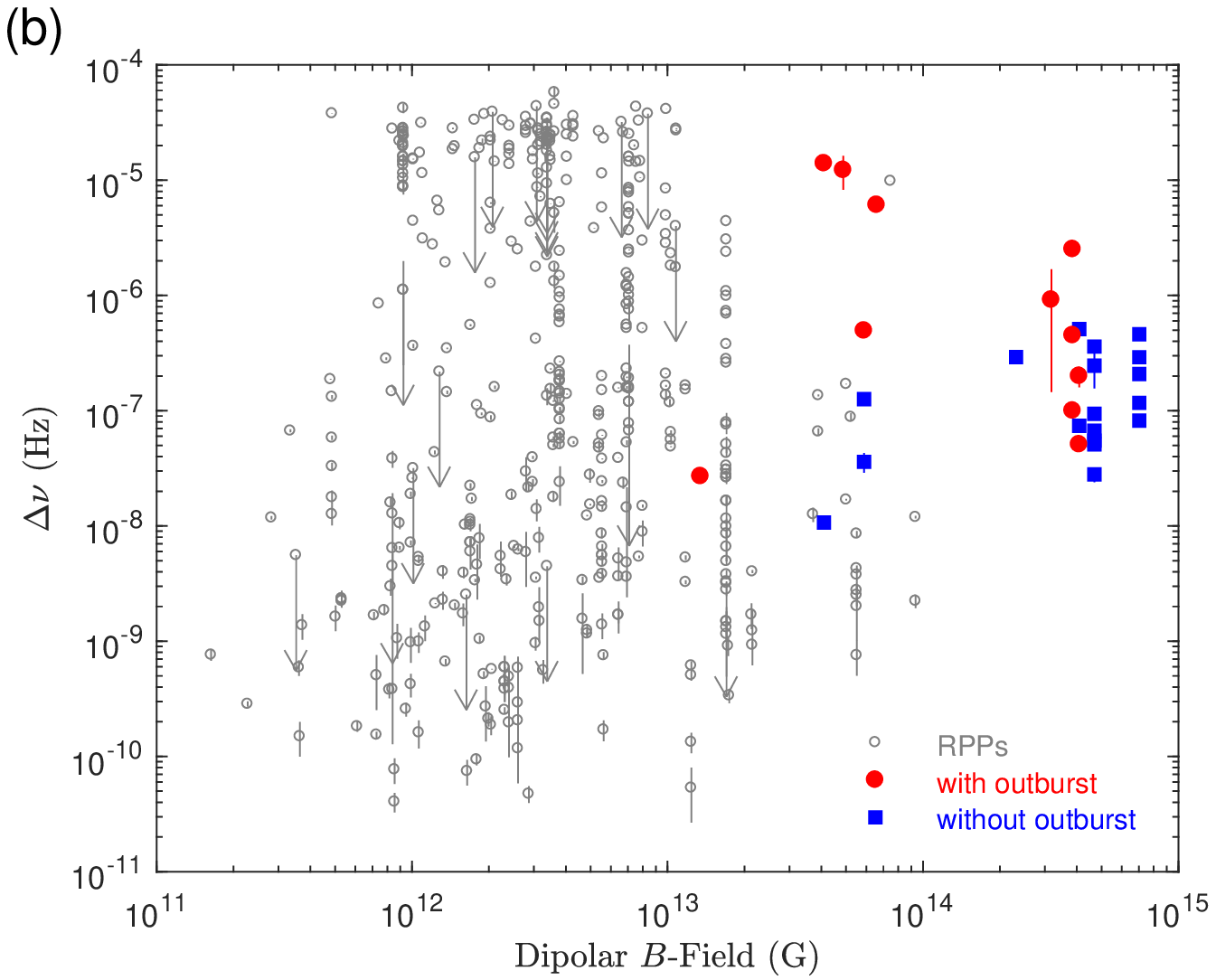}
\end{minipage}
\caption{Glitch size in $\Delta\nu$ versus (a) $\tau_c$ and (b) $B$-field for all RPPs and magnetars. Glitches in magnetars associated with outbursts are marked with red circles, while unassociated ones are marked with blue squares. Glitches in other RPPs are marked with gray open circles.}
\label{correlation_figs}
\end{figure*}

We then search for the connection between $\Delta\nu$ and the physical parameters, including characteristic age $\tau_c$ and $B$-field strength. All the glitches of canonical RPPs are also included for comparison. We adopt $\sim480$ glitches in canonical RPPs from the Pulsar Glitch Catalog\footnote{\url{http://www.jb.man.ac.uk/pulsar/glitches.html}} \citep{EspinozaLS2011}. The result is shown in Figure \ref{correlation_figs}. We found that magnetar glitches with outburst show larger size than those without outbursts, but they have no significant dependences on $\tau_c$ and $B$-field.


\begin{figure}[t]
\includegraphics[width=0.49\textwidth]{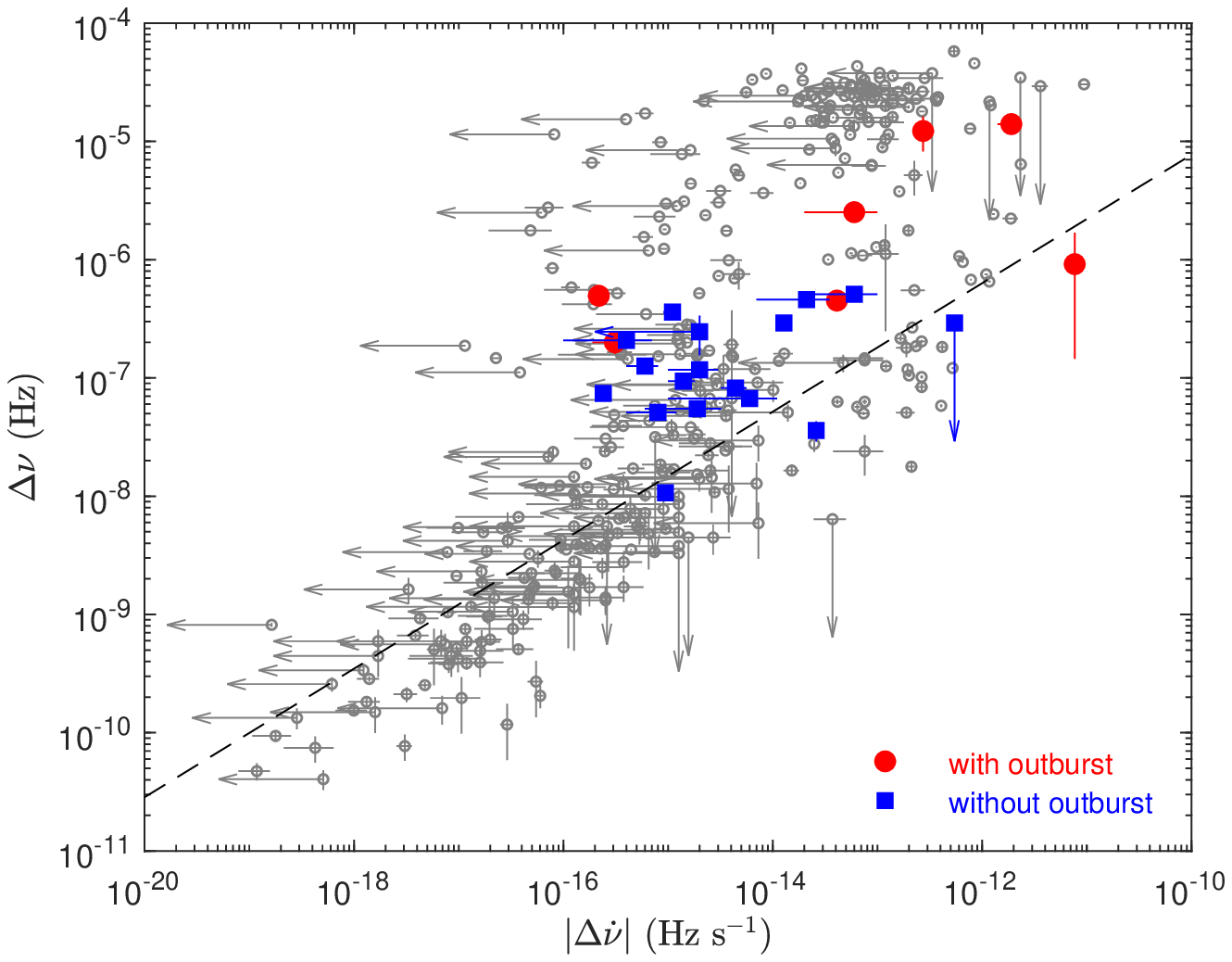}
\caption{Glitch size $\Delta\nu$ versus $\Delta\dot{\nu}$ for RPPs and magnetars. The notation is the same as those in Figure \ref{correlation_figs}. The dashed line is a fit to RPP glitches excluding Vela-like ones. \label{df_df1}}
\end{figure}

We further investigate the relation between $\Delta\nu$ and $\Delta\dot{\nu}$ (see Figure \ref{df_df1}). Since Vela-like pulsars have large $\Delta\nu\sim10^{-5}$--$10^{-4}$\,Hz and large $\Delta\dot{\nu}\sim10^{-14}$--$10^{-12}$\,Hz\,s$^{-1}$, they occupy the upper-right corner. Other glitches show a positive correlation between $\Delta\nu$ and $\Delta\dot{\nu}$, but several outliers can be seen in between ( $5\times10^{-7}$\,Hz$\lesssim\Delta\nu\lesssim10^{-5}$\,Hz and $10^{-16}$\,Hz\,s$^{-1}\lesssim\Delta\dot{\nu}\lesssim10^{-11}$\,Hz\,s$^{-1}$). Magnetars’ glitches without outburst followed the positive trend well, implying that they belong to the major glitch class. On the other hand, those glitches with outbursts are distributed between the positive correlation trend and the Vela-like glitches. They could belong to the outliers and may have different triggering mechanisms.

\begin{figure}[t]
\includegraphics[width=0.49\textwidth]{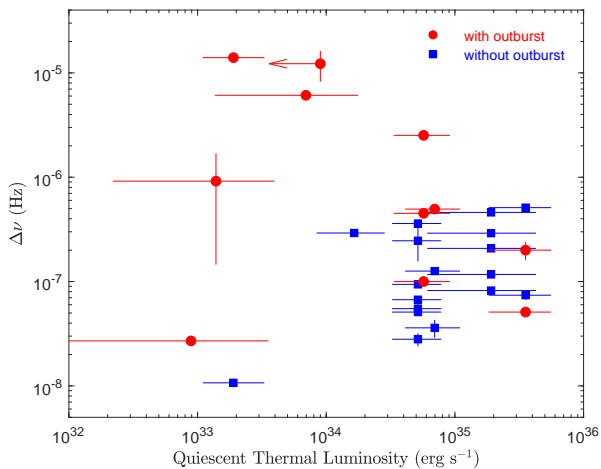}
\caption{Glitch size $\Delta\nu$ versus thermal luminosity for magnetars.\label{lum_df}}
\end{figure}

It has been proposed that magnetars have strong toroidal $B$-fields. This non-dipolar term drives the decay of the $B$-field and heats the surface of the magnetars \citep{PonsG2007,PonsMG2009,GlampedakisJS2011}. The thermal luminosity could provide hints about the hidden $B$-field components and ages of magnetars \citep{PernaP2011,ViganoRP2013}. Therefore, we plot the glitch sizes against thermal luminosities in Figure \ref{lum_df}. A more luminous magnetar is believed to be a younger one with a higher total $B$-field. Most of the glitches without outbursts are observed from these bright sources. Glitches with outbursts have a size distribution with a higher mean value and a wider deviation. Unfortunately, the timing behaviors of those transient magnetars in quiescence are difficult to monitor due to insufficient sensitivity of current X-ray observatories. PSR\,J1119$-$6127 is the only source that show one glitch without outburst ($\Delta\nu=1.1\times10^{-8}$\,Hz) and a quiescent thermal luminosity of $\sim2\times10^{33}$\,erg\,s$^{-1}$. In contrast, its glitch accompanied with outburst has a much larger size of $\Delta\nu=1.4\times10^{-5}$\,Hz. This provides a hint that the size could probably play an important factor in the triggering of outbursts.

\section{Discussion}\label{discussion}

We have collected historical glitch events of magnetars and found that glitches associated/unassociated with outbursts could have a bimodal distribution in $\Delta\nu$ although it could be biased due to a limited sample. Moreover, we do not observe significant age and $B$-field dependences of glitches in magnetars. They are consistent with glitches of canonical RPPs with $\tau_c\lesssim10^5$\,yr on the $\Delta\dot{\nu}$--$\Delta\nu$ plot. 

The current leading neutron star model suggests a superfluid layer under the solid crust. The angular momentum of the superfluid component is proportional to the density of vortices, which are pinned to the lattice in the inner crust. This effectively forms a detached component containing a higher angular momentum as the neutron star (NS) spins down. When the pinning force suddenly brakes, the vortices migrate outwards and bring angular momentum out. The superfluid component is attached to the rest of the NS and causes a sudden spin-up \citep{AndersonI1975}. The unpinning could be triggered by the Magnus force, crust rearrangement, and thermal fluctuation \citep{AlparCC1996,Middleditch2006,Melatos2009}. 

From statistics, radiative outbursts are almost always accompanied with glitches \citep{DibK2014}. Radiative outbursts are believed to be determined by the untwisting of the closed field lines \citep{BeloborodovT2007, Beloborodov2009}. The footprints of the magnetic field lines could be twisted by the motion of the crust. Hence, the triggering mechanism and location of the glitch could determine whether radiative outbursts will occur or not \citep{ArchibaldKS2017}. Moreover, the degree of twist could also correlate with the glitch size. We compared the glitch size and the flux increment of the outburst but found no significant correlation.

It was also suggested that glitches are always accompanied by radiative events but those bright sources have limited flux increases and much shorter decay time scales compared to faint magnetars \citep{PonsR2012}. Two tiny radiative outburst events accompanied with glitches were indeed observed in 4U~0142 \citep{GavriilDK2011,ArchibaldKS2017}. This could explain the lack of correlation between the glitch size and flux increment. However, a difference in size between glitches with/without outbursts is seen in Figures \ref{df_df1_histogram} and \ref{df_df1}. Those glitches with outbursts and small $\Delta\nu$ values are mainly observed from bright magnetars. Their sizes are comparable to those glitches of regular RPPs with $\tau_c<10^5$\,yr. Other glitches with huge sizes are observed in faint sources with violent and long outbursts. They are not located on the linear trend of regular RPPs and the clustering region of Vela-like pulsars (Figure \ref{df_df1}). We suggest that these glitches have a more violent deformation of crust and cause a significant twist of the $B$-field lines. Similar events could also be seen in canonical RPPs but their $B$-fields are not strong enough to trigger radiative events. 

Because the sample remains limited, we are unable to determine if violent glitches with outbursts are more often to occur in faint magnetars or high $B$-field RPPs. Fortunately, PSR~J1119$-$6127 provides a good opportunity to test the connection between glitches and outbursts because its timing behavior in the quiescent state can be achieved in the radio band \citep{JanssenS2006, WeltevredeJE2011, WeltevredeJE2011}. The glitch accompanied with outburst has the largest size, supports the above idea. Therefore, monitoring timing behaviors of high $B$-field RPPs in radio bands could play an important role to explore the connection between glitch size and the outburst behaviors. Moreover, monitoring faint magnetars in quiescence with future X-ray missions is also critical to see if there are any glitches without triggering outbursts.

\section{Summary}\label{summary}
We have carried out a comprehensive analysis of glitches in the current magnetar sample. A bimodal distribution of $\Delta\nu$ is observed. The size does not show significant correlation with $\tau_c$ and $B$-field. Glitches without outbursts are fully consistent with glitches in regular RPPs on the $\Delta\nu$--$|\Delta\dot{\nu}|$ plot, while those ones with outburst have a distribution with a larger size and they are more likely consistent with those glitches scattered between regular RPPs and Vela-like pulsars in $\Delta\nu$--$|\Delta\dot{\nu}|$ plot. Unfortunately, the lack of knowledge about the timing properties of low-luminosity magnetars in quiescence prevents us to draw a strong conclusion on the connection between timing irregularities and radiative outbursts. Monitoring them with future X-ray mission and monitoring high $B$-field RPPs in other wavelengths will be helpful for building a complete sample.

\section*{Acknowledgments}

We thank the referee for the comments that improved this paper and Prof. Kwong Sang Cheng for useful discussion. This research is supported by a \fundingAgency{GRF grant} of the \fundingAgency{Hong Kong Government} under \fundingNumber{HKU 17300215P}.


\end{document}